\documentclass[preprint,aps,tightenlines]{revtex4}
\usepackage[dvipdf]{graphicx}

\begin{document}
\title{Parallel Temperatures in Supersonic Beams: Ultra Cooling of Light Atoms Seeded in a Heavier Carrier Gas.}

\author{A. Miffre, M. Jacquey, M. B\"uchner, G. Tr\'enec and J. Vigu\'e}
\address{ Laboratoire Collisions Agr\'egats R\'eactivit\'e -IRSAMC
\\Universit\'e Paul Sabatier and CNRS UMR 5589
\\ 118, Route de Narbonne 31062 Toulouse Cedex, France
\\ e-mail:~{\tt jacques.vigue@irsamc.ups-tlse.fr}}

\date{\today}

\begin{abstract}

Supersonic expansion is a very powerful tool to produce an atomic
beam with a well defined velocity and, by seeding a test gas in
such an expansion, the energy of the test gas can be transferred,
at least partially, to the very-low-temperature carrier gas. The
case usually studied is the one of a heavy gas seeded in a light
carrier gas and, in this case, the parallel temperature of the
seeded gas is always larger than the one of the carrier gas. In
the present paper, we study the opposite case which has received
less attention: when a light gas is seeded in a heavier carrier
gas, the parallel temperature can be substantially lower for the
seeded gas than for the carrier gas. This effect has been first
observed by Campargue and coworkers in 2000, in the case of atomic
oxygen seeded in argon. In the present paper, we develop a
theoretical analysis of this effect, in the high dilution limit,
and we compare our theoretical results to several experimental
observations, including a set of measurements we have made on a
beam of lithium seeded in argon. The agreement between theory and
experiments is good.

\end{abstract}

\maketitle

%%%%%%%%%%%%%%%%%%%%%%%%%%%%%%%%%%%%%%%%%%%%%%%%%%%%%%%%%%%%%%%%%%%%%%

\section{Introduction}

We have recently observed that, in a supersonic beam of argon
seeded with lithium, the parallel temperature of lithium is
roughly one third of the parallel temperature of argon
\cite{miffre04}. A similar effect was reported in 2000 by
Campargue and co-workers \cite{lebehot01} on a beam of atomic
oxygen seeded in argon: the oxygen parallel temperature was close
to half the argon parallel temperature. This result was found to
be in agreement with a calculation developed by the same authors.

This effect is surprising because one would naively expect that
the parallel temperature of the seeded gas cannot be lower than
the parallel temperature of the carrier gas: the expansion of the
carrier gas is the source of the cooling effect and the energy of
the seeded gas is transferred by collisions to the carrier gas,
but the transfer should not be complete. This view is too naive as
shown below, but it is supported by all the experiments involving
a heavy gas is seeded in a light carrier gas: the parallel
temperature of the seeded gas always exceeds the one of the
carrier gas and the ratio of these two temperatures increases
steadily with the mass ratio, from $2.2 - 2.5$ for an He-Ar
mixture, up to $4-4.5$ for He-Xe mixture, and reaching $13-15$ for
H$_2$-Xe mixture, following results due to Campargue and
co-workers \cite{campargue81,campargue84}. At the same time, the
mean velocities of the two components are different and this
difference, the velocity slip effect \cite{anderson67,miller69},
increases also with the mass ratio.

The theory giving the terminal temperatures in supersonic beams of
a pure monoatomic gas was developed by many authors including
Anderson and Fenn \cite{anderson65}, Hamel and Willis
\cite{hamel66}, Knuth and Fisher \cite{knuth68}, Miller and Andres
\cite{miller69}, Toennies and Winkelmann \cite{toennies77} and by
Beijerinck and Verster \cite{beijerinck81}. The extension to the
case of binary mixtures was made by Cooper and Bienkowski
\cite{cooper67}, by Anderson and coworkers
\cite{anderson67,raghuraman77}, by Chesneau and Campargue
\cite{chesneau85} (the subject has been reviewed by D.R. Miller
\cite{miller88}). Because a large majority of experiments
corresponds to a heavy species seeded in a light carrier gas, the
calculations were done in this case (reference \cite{lebehot01}
being an exception). These calculations established that the
terminal parallel temperatures of the two species are not equal
\cite{raghuraman77} and were able to explain the observed ratio of
parallel temperatures of the seeded and carrier gas
\cite{cattolica79,takahashi84,chesneau85}. In the present paper,
we apply this theory to the case of mixtures of monoatomic gases.
We need several approximations to make an analytic theory and the
most important one consists in neglecting the velocity slip
effect. This approximation is good when the seeded gas atomic mass
$m_2$ is smaller than or comparable to the one of the carrier gas
$m_1$. We are able to calculate the ratio of parallel temperatures
of the seeded and carrier gases, as a function of their
interaction potentials and atomic masses.

This theory provides a physical explanation of the ultra cooling
effect: during a supersonic expansion of a pure gas, the basic
phenomenon is the geometrical cooling effect of the perpendicular
temperature, as illustrated by figure 1 of reference
\cite{toennies77}, and this cooling effect is transferred by
collisions to the parallel temperature.  When a gas is seeded in
such an expansion, with a high dilution, the parallel temperature
of the seeded gas is coupled to its own perpendicular temperature
as well as to the parallel and perpendicular temperatures of the
carrier gas. At the end of the expansion, the perpendicular
temperatures of the seeded and carrier gases keep on decreasing
and become considerably lower than the parallel temperature of the
carrier gas which is frozen. Therefore, the terminal parallel
temperature of the seeded gas can become lower than the same
quantity for the carrier gas, provided that the collisional
coupling of the temperatures acts for a longer time for the seeded
gas than for the carrier gas. We must compare the efficiency of
collisional exchange of kinetic energy during the various possible
collisions. Two effects govern this comparison: i) the relative
ranges of the seeded gas-carrier gas interaction and the
interaction between two atoms of the carrier gas; ii) the mass
ratio as it plays a very important role in the way energy is
redistributed during a collision. Among these two effects, the
mass ratio usually dominates and we predict that the parallel
temperature of the seeded gas can be considerably lower than the
parallel temperature of the carrier gas when $m_2 \ll m_1$.

The present paper is organized as follows. We first recall the
theory of the terminal temperatures in a pure gas expansion, thus
introducing the needed ideas, notations and equations. Then, we
generalize this calculation to the case of a gas mixture, in the
limit of a high dilution. In the final part, we compare the
available experiments with our calculations.

\section{Terminal parallel temperature in supersonic expansion
of a pure monoatomic gas}

In this part, we briefly recall the theory following the 1977
paper of Toennies and Winkelmann \cite{toennies77} (noted below
TW) and we introduce the simplifications used by Beijerinck and
Verster \cite{beijerinck81} in 1981 (noted below BV).

\subsection{Equations describing the cooling effect during a
supersonic expansion}

The starting point is the Boltzmann equation in the steady state
regime:

\begin{equation}
\label{m0} {\mathbf v}\cdot{\mathbf {grad}}\left[n({\mathbf
r})f({\mathbf r},{\mathbf v})\right] = \left( \frac{\partial
\left(nf\right)}{\partial t}\right)_{coll}
\end{equation}

\noindent  $n({\mathbf r})$ is the gas density and $f({\mathbf
r},{\mathbf v})$ the normalized velocity distribution. Following
TW, the supersonic expansion is approximately described, near its
axis, by a spherically symmetric flow. The velocity distribution
is assumed to remain Maxwellian, with different parallel and
perpendicular temperatures $T_{\|}$ and $T_{\bot}$:
\begin{equation}
\label{m1} f({\mathbf r},{\mathbf v}) = \left( \frac{m}{2 \pi k_B
T_{\|}}\right)^{1/2} \times \frac{m}{2 \pi k_B T_{\bot}}
\exp\left[ -\frac{m \left(v_{\|}-u\right)^2}{2k_B T_{\|}}- \frac{m
v_{\bot}^2}{2k_B T_{\bot}}\right]
\end{equation}

\noindent where $m$ is the atomic mass and $u$ is the local
hydrodynamic velocity. Then, TW express the Boltzmann equation in
spherical coordinates and apply the method of moments to obtain
four differential equations coupling the density $n$, the
hydrodynamic velocity $u$ and the temperatures $T_{\|}$ and
$T_{\bot}$ (see equations collected in table I of TW). After some
algebra and neglecting small terms proportional to $k_BT_{\|}/(m
u^2) =1/(2 S_{\|}^2)$ where $S_{\|}$ is the parallel speed ratio
$S_{\|} = u/\sqrt{2kT_{\|}/m}$ (a good approximation as $S_{\|}$
becomes large in most supersonic expansions), one obtains two
differential equations coupling the parallel and perpendicular
temperatures:

\begin{equation}
\label{m9a} \frac{d T_{\|}}{dz} = - 2 {\mathcal F}
\end{equation}

\begin{equation}
\label{m9b} \frac{d T_{\bot}}{dz} =- \frac{ 2 T_{\bot}}{z} +
{\mathcal F}
\end{equation}

\noindent where, following BV notation, we use $z$ to measure the
distance from the nozzle. ${\mathcal F}$ is a collision term given
by:

\begin{equation} \label{m10}
{\mathcal F} = \frac{n}{2k_B u} \int g \frac{d\sigma(g)}{d\Omega}
\Delta E f({\mathbf v}_1) f({\mathbf v}_2) d^3{\mathbf v}_1 d^3
{\mathbf v}_2 d\Omega
\end{equation}

\noindent In equation (\ref{m10}), ${\mathbf v}_1$ and ${\mathbf
v}_2$ are the atom velocities before the collision and ${\mathbf
g}= {\mathbf v}_1 - {\mathbf v}_2$ is their relative velocity of
modulus $g$, while $d\sigma(g)/d\Omega$ is the differential
cross-section. $\Delta E$ is the energy transferred during one
collision from the parallel degree of freedom to the perpendicular
ones for the two atoms. After averaging over the azimuth
describing the direction of the final relative velocity around the
initial relative velocity, $\Delta E$ is given by :

\begin{equation}
\label{m11}\left\langle \Delta E \right\rangle = \frac{m}{8}
\left[g_{\bot}^2 - 2 g_{\|}^2\right] \left[1 - \cos^2\chi\right]
\end{equation}

\noindent where $\chi$ is the deflection angle.

\subsection{Simplification of these equations}

BV have shown that the coupling term ${\mathcal F}$ is well
approximated by a linear function of $(T_{\|}- T_{\bot})$ given by
${\mathcal F}\approx  \Lambda(z) \left(T_{\|}- T_{\bot}\right)/2$
with $\Lambda(z) = 16n(z) \Omega^{\left(2,2\right)}(T_m)/(15
u_{\infty})$. Here, $\Omega^{\left(l,s\right)}(T)$ is a thermal
average of the collision cross-section $Q^{(l)}$, both defined in
reference \cite{hirschfelder54} (see also Appendix A). $T_m =
(T_{\|} + 2 T_{\bot})/3$ is the weighted mean of the parallel and
perpendicular temperatures. The hydrodynamic velocity $u$, in
equation (\ref{m10}), has been replaced by its terminal value
$u_{\infty}= \sqrt{5 k_B T_0/m}$ in the large $S_{\|\infty}$
limit. Neglecting quantum effects, assuming a 12-6 Lennard-Jones
potential, the $\Omega^{\left(2,2\right)}(T)$ integral is given
by:

\begin{equation}
\label{m14} \Omega^{\left(2,2\right)}(T) = 2.99 \left(\frac{2k_B
T}{m}\right)^{1/2}\left(\frac{C_6}{k_B T}\right)^{1/3}
\end{equation}

\noindent where $C_6$ is the coefficient of the attractive term of
the potential. This formula, established by BV, is valid when
$k_BT$ is small with respect to the potential well depth
$\epsilon$ (see appendix A). $n(z)$ is related to the source
density $n_0$ and temperature $T_0$ by $n(z) \approx I
/(u_{\infty} z^2)$ with the intensity $I$ of the supersonic beam
given by $I= n_0 u_{\infty} z_{ref}^2$ with $z_{ref} = 0.403
\times d$ ($d$ is the nozzle diameter):

\begin{equation} \label{m16a} \frac{d T_{\|}}{dz} = -  \Lambda(z)
\left(T_{\|} -T_{\bot}\right)
\end{equation}

\begin{equation}
\label{m16b} \frac{d T_{\bot}}{dz} =- \frac{ 2 T_{\bot}}{z} +
\frac{\Lambda(z)}{2}\left(T_{\|} -T_{\bot}\right)
\end{equation}
\noindent with $\Lambda(z)$ given by:

\begin{equation}
\label{m16c} \Lambda(z) = 3.189 \times \frac{n_0
z_{ref}^2}{u_{\infty}z^2} \left(\frac{2k_B
T_m}{m}\right)^{1/2}\left(\frac{C_6}{k_B T_m}\right)^{1/3}
\end{equation}

\subsection{Scaling and integration of these equations}

When $z$ is small, the density $n(z)$ and the collision term
${\mathcal F}$ are both large and the two temperatures remain
equal, $T_{\|}= T_{\bot} = T_m$ which verifies:

\begin{equation}
\label{m17} \frac{d T_m}{dz} =- \frac{ 4 T_m}{3z}
\end{equation}

\noindent giving $T_m \propto z^{-4/3}$. BV simplified the
equations (\ref{m16a},\ref{m16b}) by introducing reduced
temperatures, obtained by dividing the temperatures $T_{\|}$,
$T_{\bot}$, $T_m$ by $T_0$ and a reduced distance $z_r=
z/z_{ref}$:

\begin{equation}
\label{m18} \frac{d T_{\| r}}{dz_r} = -  \Xi \frac{T_{m
r}^{1/6}}{z_r^2}\left(T_{\| r} -T_{\bot r}\right)
\end{equation}

\begin{equation}
\label{m19} \frac{d T_{\bot r}}{dz_r} =- \frac{ 2 T_{\bot r}}{z_r}
+ \Xi \frac{T_{m r}^{1/6}}{2z_r^2}\left(T_{\| r}
-T_{\bot1,r}\right)
\end{equation}
\noindent All the source parameters are condensed in the quantity
$\Xi = 0.813 \times n_0 d \left(C_6/k_B T_0\right)^{1/3}$, which
can be eliminated by a further scaling due to BV, $ z_r = \zeta
\Xi^{9/11}$ and $ T_{Xr} =\tau_X \Xi^{-12/11}$ (with $X = \|$ or
$\bot$), thus providing universal equations:

\begin{equation}
\label{m21} \frac{d\tau_{\bot}}{d \zeta} = -
\frac{2\tau_{\bot}}{\zeta} + \frac{\tau_{m}^{1/6}(\tau_{\|} -
\tau_{\bot})}{2 \zeta^2}
\end{equation}

\begin{equation}
\label{m22} \frac{d\tau_{\|}}{d \zeta} = -
\frac{\tau_{m}^{1/6}(\tau_{\|} - \tau_{\bot})}{\zeta^2}
\end{equation}

\noindent These equations have been integrated numerically by BV
and, using MATLAB, we have reproduced their calculation (see
figure 1 below). When $\zeta$ is large, $\tau_{\bot}\propto
\zeta^{-1}$ while $\tau_{\|}$ tends toward a limit,
$\tau_{\|\infty} \approx 1.15$, from which we can express the
final parallel temperature as a function of source parameters:

\begin{equation}
\label{m22a} T_{\|\infty}/ T_0= 1.151\times \Xi^{-12/11}
\end{equation}

\subsection{Results and tests in the case of argon expansion}

The tradition is to give the terminal value $S_{\|\infty}$ of the
parallel speed ratio rather than the terminal parallel
temperature, with the relation $T_{\|\infty} \approx 2.5 T_0
/S_{\|\infty}^2$. We have written the results of TW and BV in the
same form to facilitate their comparison:

\begin{equation}
\label{m23} S_{\|\infty} = A \left[ n_0 d
\left(C_6/k_BT_0\right)^{1/3} \right]^{\delta}
\end{equation}

\noindent TW obtained $A= 1.413$ and $\delta= 0.53$ by fitting the
results of numerical integration of their equations written
without approximations. Following the procedure we have just
recalled, BV obtained $A = 1.313$ and $\delta = 0.545$. In the
range of practical interest, when $5<S_{\|\infty}<50$, these two
theoretical formula never differ by more than $4$\% and agree when
$S_{\|\infty}\approx 18.9$.

From a fit of experimental $S_{\|\infty}$ values for argon, BV
obtained semi-empirical values of the parameters $A = 1.782$ and
$\delta = 0.495$, using $C_6(Ar-Ar)/k_B= 4.45 \times 10^{-55}$
K.m$^6$. We have verified that this semi-empirical formula
represents very well the $S_{\|\infty}$ values for argon measured
by H. D. Meyer \cite{meyer78} and we estimate the error bar on
$S_{\|\infty}$ near $\pm 7$\%.

\section{Generalization to the case of a mixture of two monoatomic gases}

\subsection{Approximations used to solve the Boltzmann equation}

We have two Boltzmann equations, one per species, noted $i=1$ for
the carrier gas and $i=2$ for the seeded gas and the collision
terms have two parts corresponding to the two collision pairs:

\begin{equation}
\label{g0} {\mathbf v_i}\cdot{\mathbf {grad}}\left[ n_i({\mathbf
r})f_i({\mathbf r},{\mathbf v_i}) \right] = \Sigma_{j=1,2}\left(
\frac{\partial \left( n_i f_i\right)}{\partial
t}\right)_{coll}(i,j)
\end{equation}

\noindent The densities $f_i({\mathbf r},{\mathbf v_i})$ are
expressed by equation (\ref{m1}), with the atomic masses $m_i$,
the temperatures $T_{\| i}$ and $T_{\bot i}$. We assume that the
hydrodynamic velocity $u$ is the same for both species. When a
heavy species is seeded in a light gas, the velocity slip effect
\cite{anderson67,miller69} is not negligible and this
approximation would be bad, but we are interested in the opposite
case, when the atomic mass of the seeded gas is smaller than or
comparable to the atomic mass of the carrier gas. When the
expansion is well in the supersonic regime, the difference of the
mean velocities is negligible with respect to the thermal velocity
in the moving frame (see below).

We consider the high dilution limit, when the seeded gas density
$n_2$ is considerably smaller than the carrier gas density $n_1$.
Because $n_2 \ll n_1$, in the Boltzmann equation for the carrier
gas distribution function $f_1$, we can neglect the $1-2$
collision term. The expansion of the carrier gas is not modified
by the seeded gas and the equations written above can be used to
calculate the parallel and perpendicular temperatures of the
carrier gas. We must simply introduce in $\Lambda(z)$ the relevant
collision integral $\Omega_{1,1}^{\left(2,2\right)}(T)$, where the
indices designate the colliding atom pair. In the Boltzmann
equation for the seeded gas $f_2$, we consider only the effect of
$1-2$ collisions, i.e. collisions with atoms of the carrier gas
and we neglect the collisions involving two atoms of species $2$.
Then, we get the following equations for the temperatures of
species $2$:

\begin{equation}
\label{g1} \frac{d T_{\|2}}{dr} = 2 {\mathcal F}_{\|,2}
\end{equation}

\begin{equation}
\label{g2} \frac{d T_{\bot2}}{dr} =- \frac{ 2 T_{\bot2}}{r} +
{\mathcal F}_{\bot2}
\end{equation}

\noindent with the collisional energy transfer terms given by:

\begin{equation} \label{g3}
{\mathcal F}_{X,2}= \frac{n_1}{k_B u} \int g
\frac{d\sigma_{1,2}(g)}{d\Omega} \Delta E_{X,2} f_1({\mathbf v}_1)
f_2({\mathbf v}_2) d^3{\mathbf v}_1 d^3 {\mathbf v}_2 d\Omega
\end{equation}

\noindent where $\Delta E_{X,2}$ ($X = \|$ or $\bot$) measures the
parallel or perpendicular energy gained by atom $2$ during the
collision with atom $1$. We can express $\Delta E_{X,2}$ with the
parallel and perpendicular components of the center of mass
velocity and of the relative velocity. However, when the four
temperatures are not equal, the product $f_1({\mathbf v}_1)
f_2({\mathbf v}_2)$ has not a simple form when expressed with
these velocities and we have not been able to calculate exactly
the resulting integrals. Therefore, to get an analytic result for
${\mathcal F}_{X,2}$, we have used the following approximation.
When the collision energy is small, the cross-sections
$Q_{1,2}^{(l)}(g)$ behave like $g^{-2/3}$ and the products
$gQ_{1,2}^{(l)}(g)$ vary slowly with $g$, like $g^{1/3}$. We take
these products $g d\sigma_{1,2}(g)/d\Omega$ out of the integrals
over ${\mathbf v}_1$ and ${\mathbf v}_2$ to get:

\begin{equation} \label{g4}
{\mathcal F}_{X,2}\approx \frac{n_1}{k_B u_{\infty}}  \int \langle
g \frac{d\sigma_{1,2}(g)}{d\Omega}\rangle d\Omega \int \Delta
E_{X,2} f_1({\mathbf v}_1) f_2({\mathbf v}_2) d^3{\mathbf v}_1 d^3
{\mathbf v}_2
\end{equation}

\noindent The calculation of this integral is described in
appendix B.

\subsection{Coupled equations describing the temperatures of the
two gases}

The coupled equations for the carrier gas $i=1$ have been
established in part II (equations (\ref{m16a}) and (\ref{m16b}))
where $T_{X}$ must be replaced by $T_{X1}$. In the differential
equations for the temperatures of the seeded gas $i=2$, we have
quantities similar to $\Lambda(z)$ but involving the
$\Omega_{1,2}^{(l,2)}(T_m)$ integrals with $l=1$ and $2$. By
introducing two dimensionless ratios $\rho_s$ and $\rho_o$, we can
express these quantities as a function of
$\Omega_{1,1}^{(2,2)}(T_m)$. $\rho_s$ is the ratio of
$\Omega_{i,j}^{\left(2,2\right)}$ collision integrals differing by
the species of the second collision partner and its value is
deduced from equation ({\ref{m14}):

\begin{equation}
\label{g8} \rho_s =
\frac{\Omega_{1,2}^{\left(2,2\right)}}{\Omega_{1,1}^{\left(2,2\right)}}
= \left[\frac{C_6(1,2)}{C_6(1,1)}\right]^{1/3} \times
\left[\frac{m_1 + m_2}{2 m_2}\right]^{1/2}
\end{equation}

\noindent $\rho_s$ depends slowly on the $C_6$ ratio and more
rapidly on the mass ratio. $\rho_o$ is the ratio of angle-averaged
cross-sections of orders $l=1$ and $l=2$, defined by $\rho_o =
\Omega_{1,2}^{\left(1,2\right)}/\Omega_{1,2}^{\left(2,2\right)} =
1.32$ (see Appendix A and our previous paper \cite{miffre04}).

\begin{eqnarray}
\label{g10} \frac{dT_{\bot2}}{dz} &=& -\frac{2 T_{\bot2}}{z} +
\Lambda(z)\rho_s \frac{m_1}{M} \left[T_{\|,av} -
T_{\bot,av}\right] \nonumber \\ &-&  4\Lambda(z)\rho_s \rho_o
\frac{\mu}{M}\left[T_{\bot2} -T_{\bot1}\right]
\end{eqnarray}
\begin{eqnarray}
\label{g10a} \frac{dT_{\|2}}{dz} &=& - 2 \Lambda(z)\rho_s
\frac{m_1}{M} \left[T_{\|,av} - T_{\bot,av}\right] \nonumber \\
&-& 4 \Lambda(z) \rho_s \rho_o \frac{\mu}{M}\left[T_{\|2}
-T_{\|1}\right]
\end{eqnarray}

\noindent $\Lambda(z)$ being given by equation (\ref{m16c}) and
$T_{X,av}$ being defined by:

\begin{equation}
\label{g11}T_{X,av} = \beta T_{X1} + \alpha T_{X2}
\end{equation}

\noindent with $\alpha=m_1/M$, $\beta = m_2/M$, $M= m_1 + m_2$ and
$\mu = m_1m_2/(m_1+m_2)$. One should remark that in $T_{X,av}$,
the weight of $T_{X1}$ is $\beta = m_2/M$ and the weight of
$T_{X2}$ is $\alpha=m_1/M$. We proceed as in part I.C and we get
two equations verified by $\tau_{\|2}$ and $\tau_{\bot2}$:

\begin{eqnarray}
\label{g13} \frac{d\tau_{\bot2}}{d \zeta} &=& -
\frac{2\tau_{\bot2}}{\zeta} + \frac{\rho_s
\tau_{m}^{1/6}}{\zeta^2} \times \frac{m_1}{M} \left(\tau_{\|,av}
-\tau_{\bot,av}\right)\nonumber \\ &-& \frac{4 \rho_s \rho_o
\tau_{m}^{1/6}}{\zeta^2}\times \frac{\mu}{M}\left(\tau_{\bot 2} -
\tau_{\bot1}\right)
\end{eqnarray}

\begin{eqnarray}
\label{g14} \frac{d\tau_{\|2}}{d \zeta} &=& - \frac{2\rho_s
\tau_{m}^{1/6}}{\zeta^2} \times \frac{m_1}{M} \left(\tau_{\|,av}
-\tau_{\bot,av}\right) \nonumber \\ &-& \frac{4 \rho_s \rho_o
\tau_{m}^{1/6}}{\zeta^2}\times \frac{\mu}{M}\left(\tau_{\| 2} -
\tau_{\|1}\right)
\end{eqnarray}

\noindent A simple test of the coherence of our calculations is to
consider that the two species have the same masses ($m_1 = m_2$)
and the same collision cross-sections ($\rho_s=1$). We then expect
that the parallel and perpendicular temperatures are independent
of the species ($\tau_{X1} = \tau_{X2}$ for all $\zeta$ values)
and this property is well verified.

Using MATLAB, we have integrated the equations
(\ref{g13},\ref{g14}) and the results are represented in figure 1.
At the end of the expansion, the parallel and perpendicular
temperatures of the seeded gas are intermediate between the same
quantities for the carrier gas. In particular, the terminal value
of the parallel temperature ratio
$\tau_{\|2,\infty}/\tau_{\|1,\infty} =
T_{\|2,\infty}/T_{\|1,\infty}$ can be substantially lower than
$1$. The absolute minimum value for this temperature ratio is
reached when the right-hand side of equation (\ref{g10a})
vanishes. Assuming that the perpendicular temperatures are both
negligible, we get the minimum possible value of the ratio
$T_{\|2,\infty}/T_{\|1,\infty}$:

\begin{equation}
\label{g15} {\mbox {Min}} \left(\frac{T_{\|2,\infty}}
{T_{\|1,\infty}}\right) = \frac{m_2 \rho_o}{m_1 + 2 m_2 \rho_o}
\end{equation}

\noindent This limiting value would be reached if the ratio
$\rho_s$ tends toward infinity. We have plotted in figure 2 the
terminal value of the ratio $T_{\|2,\infty}/T_{\|1,\infty}$ for
various values of the ratio $\rho_s$, as a function of the mass
ratio $m_2/m_1$. When the mass ratio $m_2/m_1$ is very small, the
parallel temperature of the seeded gas can be considerably smaller
than the one of the carrier gas. We can make some surprising
predictions:

\begin{itemize}
\item In the case of molecular hydrogen H$_2$ seeded in argon,
using to calculate the Ar-H$_2$ $C_6$ coefficient the combination
rule and the data of Kramer and Herschbach \cite{kramer70}, we get
$\rho_s = 2.48$ from which we predict a temperature ratio
$T_{\|2,\infty}/T_{\|1,\infty} = 0.16$, a quite large effect!

\item for a beam of potassium seeded in argon, the two atomic
masses are almost equal and, with the $C_6$ from reference
\cite{standard85}, we get $\rho_s = 1.67$. We predict a
temperature ratio $T_{\|2,\infty}/T_{\|1,\infty} = 0.77$, already
smaller than $1$.

\item for a beam of sodium seeded in neon, we calculate in the
same way $\rho_s = 1.85$, (the large $C_6$ ratio overcompensates
the effect of the mass ratio $m_2/m_1= 1.14$) and we predict a
temperature ratio $T_{\|2,\infty}/T_{\|1,\infty} = 0.77$, also
smaller than $1$. This is not a large effect, but this result,
with $T_{\|2,\infty}<T_{\|1,\infty}$ while $m_2>m_1$, is contrary
to the common belief recalled in the introduction.

\end{itemize}

\subsection{Discussion}

In addition to the approximations already done by Beijerinck and
Verster, we have neglected the velocity slip, which has been
studied in particular by Anderson \cite{anderson67} and by Miller
and Andres \cite{miller69}. Equations (2.37,2.38) of Miller
\cite{miller88}) give the velocity difference
$u_{2,\infty}-u_{1,\infty}$, which can be written with our
notations and in the high dilution limit:

\begin{equation}
\label{g16}  \frac{u_{2,\infty}-u_{1,\infty}}{u_{1,\infty}}
\approx 0.59 \left[ \frac{\sqrt{\mu m_1}}{|m_1 - m_2|} n_0 d
\left(C_6/k_BT_0\right)^{1/3} \right]^{-1.07}
\end{equation}

\noindent The $\Omega_{1,2}^{(1,1)}$ integral, which appears in
these equations, is evaluated in Appendix A. The physically
important quantity is the ratio of this velocity difference
$\left(u_{2,\infty}-u_{1,\infty}\right)$ divided by the parallel
thermal velocity of atom $1$, namely $\sqrt{2k_BT_{\|1}/m_1}$.
Using equation (\ref{m23}) in the TW form and approximating the
$1.07$ exponent by $2\times0.53$, we get:

\begin{equation}
\label{g17}
\frac{u_{2,\infty}-u_{1,\infty}}{\sqrt{2k_BT_{\|1}/m_1}} \approx
\frac{0.84}{S_{\|1\infty}} \left[
\frac{(m_1+m_2)(m_1-m_2)^2}{m_1^2 m_2}\right]^{0.53}
\end{equation}
\noindent If the terminal parallel speed ratio $S_{\|1\infty}$ is
large, the velocity slip is a small fraction of the parallel
thermal velocity of atom $1$, provided that the masses $m_1$ and
$m_2$ are not extremely different. In this case, neglecting the
velocity slip is an excellent approximation.

In equation (\ref{g3}), we have also treated the products
$gQ_{1,2}^{(l)}(g)$ as constant. Takahashi and Teshima were able
to calculate numerically these integrals when they studied the
case of a heavy gas seeded in a light gas \cite{takahashi84}.
Using their technique, it would be possible to test the accuracy
of our approximation.

Monte Carlo simulation can also be used to simulate flows of gases
and gas mixtures. The Direct Simulation Monte Carlo method
described by G. A. Bird in his book \cite{bird94} is the best
method and it has been used very early to simulate supersonic
expansions \cite{bird70,cattolica79}. This method requires much
computation in the high dilution case \cite{skovordko04} and P. A.
Skovorodko has used the Test Particle Monte Carlo method to
simulate the expansion of gas mixtures. This method also involves
an approximation, the use of Maxwell type interaction potential.
He has simulated our lithium seeded in argon expansion and his
results support our calculations, at least at the qualitative
level \cite{skovorodko04a}. In his simulations, P. A. Skovorodko
has found that the distribution of the perpendicular velocity of
lithium differs very much from a Maxwellian distribution and is
almost perfectly exponential. The fact that the velocity
distribution of the perpendicular degree of freedom cannot be
Maxwellian was analyzed in detail by Beijerinck and co-workers who
suggested to call blistering this effect
\cite{beijerinck81,beijerinck83}. This effect cannot be taken into
account by the theory described in the present paper as the
assumption of an elliptic Maxwellian velocity distribution is the
starting point of the solution of Boltzmann equation.

\section{Comparison with experimental results}

In this part, we first describe briefly the previous experimental
cases in which the light gas parallel temperature has been found
lower than the carrier gas parallel temperature. Then, we describe
our experiment with a beam of lithium seeded in argon and a new
set of temperature measurements.

\subsection{Sodium in seeded in argon}

A beam of sodium seeded in argon was built by D. Pritchard and
coworkers \cite{schmiedmayer97} with a source temperature near
$1000$ K, an argon pressure up to $p_0 =3$ bars and a nozzle
diameter $d = 70$ $\mu$m. In a particular experiment
\cite{ekstrom95}, the mean velocity was measured, $u= 1040 \pm 2$
m/s corresponding to $T_0= 1039$ K, and the rms velocity width,
deduced from the atomic diffraction pattern, was found equal to
$3.7 \pm 0.4$\% , corresponding to $T_{\|2\infty} = 4.1 \pm 0.9$
K.  Assuming that this experiment was made with the largest
pressure quoted in \cite{schmiedmayer97}, we calculate the argon
parallel temperature $T_{\|1} = 7.7\pm 1.1$ K, from which we
deduce the parallel temperature ratio $T_{\|2\infty}/T_{\|1\infty}
= 0.53 \pm 0.19$. This value is in good agreement with our
theoretical result is $T_{\|2\infty}/T_{\|1\infty} = 0.66$,
deduced from the value of $\rho_s = 1.67$ (obtained with the $C_6$
values from reference \cite{standard85}).

\subsection{Atomic oxygen seeded in argon}

In 2000, Campargue and co-workers \cite{lebehot01} operated and
characterized a beam of atomic oxygen seeded in argon produced by
a laser plasma source. The oxygen parallel temperature was roughly
half the argon parallel temperature (see figure 10 of reference
\cite{lebehot01}). The numerical model described in this paper
supports the experimental results. Let us compare this result with
our theory. Using the combination rule of Kramer and Herschbach
\cite{kramer70} and data for oxygen from reference
\cite{cummings75}, we calculate the oxygen-argon $C_6$ coefficient
$C_6(O-Ar)= 34$ atomic units and $\rho_s = 1.06$. Our theory then
predicts a temperature ratio $T_{\|2,\infty}/T_{\|1,\infty}
=0.76$, substantially larger than the experimental value near
$0.5$. This discrepancy is probably due to the fact that, because
of the very high source temperature close to $10^4$ K, the
terminal parallel temperatures are considerably higher in this
case than for usual beam experiments, in the $70-130$ K range for
oxygen and $220-150$ K range for argon. For such terminal
temperatures, we cannot expect the low temperature approximations
of the $\Omega^{\left(l,s\right)}(T)$ integrals to be valid.

\subsection{Lithium seeded in argon}

\subsubsection{The beam source}

Our beam \cite{delhuille02} is inspired by the design used by
Broyer, Dugourd and co-workers to produce lithium clusters
\cite{blanc92}. The temperatures used in our experiment are
usually equal to $973$ K for the back part of the oven,
corresponding to a lithium vapor pressure of $0.55$ millibar and
to $1073$ K for the front part. The nozzle is a hole of $200$
$\mu$m diameter drilled in a stainless steel $0.3$ mm thick wall.
The argon gas (from Air Liquide, $99.999$ \% stated purity) is
further purified by a purifying cartridge also from Air Liquide
and its pressure can be varied from $150$ to $800$ millibar,
limited by the throughput of our oil diffusion pump (Varian VHS400
with a $8000$ l/s pumping speed) backed by a Leybold D65B roughing
pump ($65$ m$^3$/hour pumping speed). For a source pressure $p_0=
300$ millibar, the pressure at the diffusion pump is $8 \times
10^{-4}$ millibar. We use a skimmer from Beam Dynamics with $0.97$
mm aperture at a $20$ mm distance from the nozzle. After the
skimmer, the lithium beam is in a separate vacuum tank pumped by
an oil diffusion pump (Varian VHS6 with a $2400$ l/s pumping
speed) fitted with a water cooled baffle. Under beam operation,
when $p_0= 300$ millibar, the pressure in this chamber is $3
\times10^{-6}$ millibar.

\subsubsection{Doppler measurement of the parallel and
perpendicular velocity distribution}

In the center of the second vacuum tank, i.e. $225$ mm after the
skimmer, the lithium beam is crossed by two laser beams $A$ and
$B$. The angle between the atomic beam and the laser beams are
$\theta_A =47.9\pm 0.5^\circ$ and $\theta_B\approx 90^\circ$. The
first order Doppler effect is sensitive only to the projection
$v_p$ of the velocity on the laser beam axis. For a laser beam
making the angle $\theta$ with the axis of the atomic beam, the
distribution of $v_p$ is deduced from equation (\ref{m1}):
\begin{equation}
\label{exp1} f(v_p) =\left( \frac{m}{2 \pi k_B
T(\theta)}\right)^{1/2} \exp\left[ -\frac{m \left(v_p
-u\cos\theta\right)^2}{2k_B T(\theta)}\right]
\end{equation}with $T(\theta) = T_{\|}\cos^2\theta +
T_{\bot}\sin^2\theta$. The distribution $f(v_p)$ is centered at
$v_p=u \cos\theta$ and its width is characterized by a weighted
mean of the parallel and perpendicular temperatures. The
fluorescence intensity as a function of the laser frequency
reflects the velocity distribution if the natural width of the
excited transition is negligible with respect to the Doppler width
and if saturation broadening of the transition as well as laser
frequency jitter are both negligible.

The laser excites successively the hyperfine components of the
$^2S_{1/2}$ - $^2P_{3/2}$ resonance transition of lithium at $671$
nm \cite{sansonetti95}. The very small hyperfine splittings of the
upper state can be neglected. The ground state has two hyperfine
components $F=1$ and $F=2$, with a splitting equal to $803.5$ MHz.
The natural width of the transition is $\Gamma/2\pi = 5.87$ MHz
\cite{mcalexander96} and we use laser power density of the order
of $10^{-2}$ mW/cm$^2$, corresponding to a saturation parameter
$s\approx 5\times10^{-3}$ and a negligible broadening of the
excitation line. Our single frequency cw dye laser pumped by an
argon ion laser has a linewidth of the order of $1$ MHz, thanks to
the H\"ansch-Couillaud \cite{hansch80} frequency stabilization
technique.

Figure 3 shows such a laser induced fluorescence spectrum: for
beam $A$, the Doppler full widths are of the order of $200$ MHz
and the natural width as well as the laser linewidth are
negligible. During our experiment, we have also recorded a
saturated absorption spectrum in a heat pipe oven: this signal
provides Doppler-free peaks which are useful if one wants to
measure accurately the mean velocity $u$ of the atomic beam,
because, in some experiments, the laser beam $B$ is not exactly
perpendicular to the atomic beam.

\subsubsection{Experimental results}

From the analysis of the fluorescence signals, we can deduce the
beam mean velocity $u= 1010 \pm 10$ m/s, slightly less than the
value deduced from the source temperature, $u_{\infty}= \sqrt{5
k_B T_0/m} = 1056$ m/s. The width of the $B$ peaks is related to
the perpendicular temperature of the beam but we think that the
perpendicular temperature $T_{\bot}$ thus deduced is overestimated
because our detector is not observing only one streamline. The
width of the $A$ peaks provide the weighted mean $T(\theta)$ of
the temperatures, from which we can deduce the parallel
temperature. As we think that our measurement may overestimate the
perpendicular temperature, we have neglected the contribution of
the perpendicular temperature when extracting the parallel
temperature from $T(\theta)$, i.e. we have used:
\begin{equation}
\label{exp3}  T_{\|} =T(\theta) / \cos^2\theta
\end{equation} so that we get an upper limit of the parallel temperature
$T_{\|}$. In table I, we have also given the fluorescence signal
intensity which gives an idea of the intensity of the lithium
beam. This intensity decreases when the argon pressure increases.
This behavior is probably due to the interaction of the beam with
the residual gas near the skimmer.

\begin{table}
  \begin{center}
  \begin{tabular}{|c|c|c|c|}
  \hline
  Source pressure (mbar)   &  $T_{\|2}(K)$  &  $T_{\bot2}(mK)$  & Fluorescence signal(a.u.) \\
  \hline
   200 & 10.1 & 493 & 8.5 \\
  \hline
   267 &  7.6 & 514 & 7.9 \\
  \hline
   333 &  6.6 & 497 & 7.0 \\
  \hline
   400 &  6.0 & 488 & 5.8 \\
  \hline
   467 &  6.1 & 546 & 6.6 \\
  \hline
   534 &  6.0 & 593 & 5.0 \\
  \hline
   600 &  5.3 & 550 & 5.6 \\
  \hline
  \end{tabular}
   \end{center}
 \end{table}

The measured parallel temperature is plotted as a function of
argon pressure in figure 4. The full curve is a plot of our
results with the ratio $T_{\|2}/T_{\|1}$ taken equal to its
predicted value $0.38$. The dashed curve is obtained by fitting
the value of this ratio $T_{\|2}/T_{\|1}$ to the four data points
with an argon pressures smaller than 400 millibar: the fitted
value is equal to $0.31$. $T_{\|1}$ is related to the source
parameters by equation (\ref{m23}), using the semi-empirical
coefficients of BV. For larger pressures, the parallel temperature
remains constant while theory predicts that it should still
decrease. We think that this discrepancy may be due to the
interaction of the beam with the residual gas discussed above.

Another explanation could be cluster formation in the expansion.
Following the results of Hagena \cite{hagena74} further discussed
by Beijerinck and Verster \cite{beijerinck81}, for an argon beam,
a safe upper limit to prevent cluster formation is to keep the
quantity $p_0 d^{0.88} T_0^{-2.3} < 4.9 \times 10^{-7}$ mbar
m$^{0.88}$ K$^{-2.3}$. At our largest source pressure $p_0 = 600$
mbar, the quantity $p_0 d^{0.88} T_0^{-2.3} = 3.6\times 10^{-8}$
mbar m$^{0.88}$ K$^{-2.3}$ is well below this upper limit. As the
LiAr molecule is considerably less bound than the Ar$_2$ molecule
\cite{aziz93,bruhl95} and as the lithium concentration is very
small, we think that the threshold for the formation of mixed
clusters is also not reached.

\section{Summary}

The knowledge concerning the terminal parallel temperatures
produced by supersonic expansions of a mixture of monoatomic gases
has been developed for many years and the case of heavy atom
seeded in a light carrier gas, which has been subject of almost
all the studies, is extremely well understood. The opposite case
(a light gas seeded in an heavier carrier gas) had been rarely
studied, an exception being the work of Campargue and co-workers
\cite{lebehot01} in 2000. This opposite case might seem of limited
interest but some experiments (for instance, atom interferometry)
require a rather slow atomic beam with a small velocity
dispersion: in this case, seeding a light gas in an heavier
carrier gas is the most natural solution.

In the present paper, we have developed a theoretical analysis by
making a slight extension of the theoretical works of Toennies and
Winkelmann \cite{toennies77} and of Beijerinck and Verster
\cite{beijerinck81}. With some new approximations, we obtain an
analytic treatment of the problem, valid in the high dilution
limit. We thus rationalize the surprising experimental result we
had observed after Campargue and co-workers \cite{lebehot01}: the
parallel temperature of the seeded gas can be lower than the same
quantity for the carrier gas. We have found that this effect
depends on two quantities, the ratio $\rho_s$ of cross-sections
for the two gases, defined by equation (\ref{g8}), and the ratio
of the atomic masses $m_2/m_1$. Finally, the physical basis of
this ultra cooling effect is not mysterious and we have explained
it in the introduction.

In a final part, we have discussed the available experimental
evidences of this effect, due to other groups
\cite{lebehot01,schmiedmayer97,ekstrom95} and we have described
our experiment with a beam of lithium seeded in argon. We have
presented a set of experimental data considerably larger than in
our first study \cite{miffre04}. The agreement between our
theoretical prediction and experiments can be considered as good
in the cases of sodium and lithium seeded in argon. The agreement
is less good for the argon-oxygen beam but we understand why it is
so.

We think that the ultra cooling effect described here is now well
established and some further studies would be very interesting.
First of all, it should be easy to observe this effect in several
other experimental cases, some of them being discussed in part
III.B: a particularly striking case would be to seed molecular
hydrogen in argon or even heavier rare gases. From the theoretical
point of view, it is interesting to investigate this effect by
other methods and Monte Carlo simulation, which has been already
applied to this problem by P. A. Skovorodko \cite{skovorodko04a},
is particularly well suited as the approximations needed to solve
Boltzmann equation can be removed.

\section{Acknowledgements}

We thank J.P. Toennies, H. C. W. Beijerinck, U. Buck and P. A.
Skovorodko for helpful discussions. We also thank Ph. Dugourd and
M. Broyer for advice concerning the design of the oven of our
lithium supersonic beam, J. Schmiedmayer for advice concerning the
use of LD688 dye as well as R. Delhuille, L. Jozefowski and C.
Champenois for their important contributions to the construction
of our experimental setup. We thank CNRS-SPM, R\'egion Midi
Pyr\'en\'ees, IRSAMC and Universit\'e P. Sabatier for financial
support.

\section{Appendix A: Weighted cross-sections; their definition and
calculation of their temperature dependence}

\subsection{Definitions}

We consider the collision between two atoms of species $i$ and $j$
with a relative velocity $g$ and a reduced mass $\mu =
m_im_j/(m_i+m_j)$. The definitions and many results are taken from
the book Molecular theory of gases and liquids, by Hirschfelder,
Curtiss and Bird \cite{hirschfelder54}. Starting from the
differential cross-section $d\sigma_{i,j}(g)/d\Omega$, the
angle-weighted cross-sections $Q_{i,j}^{(l)}(g)$ are defined by
(equation 8.2-2 of \cite{hirschfelder54}) :
\begin{equation}
\label{A1} Q_{i,j}^{(l)}(g) = \int\frac{d\sigma_{i,j}(g)}
{d\Omega} (1- \cos^l\chi) d\Omega
\end{equation}
\noindent where $\chi$ is the deflection angle. In classical
mechanics, the deflection angle $\chi$ is a function of the impact
parameter $b$ and the relative velocity $g$. $Q_{i,j}^{(l)}(g)$ is
then given by:

\begin{equation}
\label{A2} Q_{i,j}^{(l)}(g) = \int (1- \cos^l\chi)2\pi b d b
\end{equation}

\noindent The thermal averages
$\Omega_{i,j}^{\left(l,s\right)}(T)$ are given by equation (8.2-3)
of reference \cite{hirschfelder54}:
\begin{equation}
\label{A3} \Omega_{i,j}^{\left(l,s\right)}(T) =
\sqrt{k_BT/(2\pi\mu)} \int_0^{\infty} Q_{i,j}^{(l)}(g)
\gamma^{2s+3} \exp(-\gamma^2) d\gamma
\end{equation}
\noindent with $\gamma^2 = \mu g^2/(2k_BT)$. Three circumstances
contribute to make classical mechanics a good approximation in the
experimental cases considered here:

\begin{itemize}
\item  the quantum character of the interaction between two identical atoms
of mass $m$ is measured by the quantum parameter $\eta =
\hbar^2/(m\sigma^2 \epsilon)$ introduced by Stwalley and Nosanow
\cite{stwalley76} (see also De Boer and Lunbeck \cite{deboer48})
where $\epsilon$ is the potential well depth and $\sigma$ the core
radius. A large $\eta$ value indicates a highly quantum behavior
while a small one corresponds to a quasi-classical behavior.
$^4$He dimer is highly quantum with $\eta \approx 0.18$
\cite{stwalley76}, while $Ar$ dimer \cite{aziz93} with
$\eta\approx 0.75 \times10^{-3}$ and $^7Li-Ar$ \cite{bruhl95} with
$\eta\approx 6.6 \times 10^{-3}$ have a weak quantum character.

\item the differential cross-section presents several types of
quantum effects. These effects are largely reduced by the angle
average following equation (\ref{A1}) and further reduced by the
thermal average following equation (\ref{A3}).

\item a quantum effect which survives these averages is the
behavior of the cross-section at very low energy. This is the
quantum threshold regime, which extends up to an energy of the
order of $\hbar^3 \mu^{-3/2}C_6^{-1/2}$ for an potential with a
$C_6/r^6$ long range, following Julienne and Mies
\cite{julienne89}. This effect explains the very large
helium-helium cross-section at low energy \cite{toennies77}. For
argon-argon and lithium-argon collisions, we find that this regime
could be observed for temperatures of the order of $10^{-2}$ and
$2\times 10^{-2}$ Kelvin respectively, well below the lowest
temperatures obtained in argon expansions, of the order of $1$
Kelvin \cite{beijerinck81,meyer78}.
\end{itemize}

\subsection{Calculations with a Lennard-Jones $12-6$ potential}
We consider an interaction potential of the $12-6$ Lennard-Jones
type:
\begin{equation}
\label{A5} V(r) = \frac{C_{12}}{r^{12}} - \frac{C_{6}}{r^{6}}
=4\epsilon \left[\frac{\sigma^{12}}{r^{12}}
-\frac{\sigma^{6}}{r^{6}}\right]
\end{equation}
\noindent using the usual reduced quantities (see paragraph 8.2 of
\cite{hirschfelder54}). In a first step, we have calculated the
deflection function $\chi(b)$ as a function of the reduced energy
$E^{\ast} = \mu g^2/(2\epsilon)$, by numerical integration. Our
calculation reproduces well the analytic deflection function of BV
for $E^{\ast}\leq 0.1$ (see appendix A of \cite{beijerinck81}).

We have then calculated the $Q^{(l)}(g)$ cross-sections with $l=1$
and $l=2$ as a function of $E^{\ast}$. In the low-energy range,
$E^{\ast}\ll 1$, these two cross-sections are well approximated by
an $(E^{\ast})^{-1/3}$ behavior and, to make this behavior very
clear, we have plotted in figure 5 the variation of $Q^{(l)}(g)
E^{\ast 1/3}$ as a function of $E^{\ast}$. At low energy, our
calculation presents a small numerical noise associated to the
orbiting singularity. We have not tried to reduce this noise by
introducing some analytic approximations of the function $\chi(b)$
near this singularity.

When $E^{\ast}\longrightarrow 0 $, we get:

\begin{equation}
\label{A51} Q^{(1)}(g) \approx (7.81 \pm 0.01) \sigma^2 E^{\ast
-1/3}
\end{equation}
\noindent and

\begin{equation}
\label{A52} Q^{(2)}(g) \approx (5.95 \pm 0.01) \sigma^2 E^{\ast
-1/3}
\end{equation}

Then, we have calculated numerically the $\Omega^{\left(l,s
\right)}$ integrals as a function of the reduced temperature
$T^{\ast}= k_BT /\epsilon$ and we have verified their low energy
limits analytically (this is easy as the integral in equation
(\ref{A3}) can be expressed by a Gamma function). The various
$\Omega^{\left(l,s\right)}(T)$ take a form analogous to equation
(\ref{m14}):
\begin{equation}
\label{A6} \Omega^{\left(2,2\right)}(T) = C^{(l,s)}(T^{\ast})
\left(\frac{k_B T}{\mu}\right)^{1/2}\left(\frac{C_6}{k_B
T}\right)^{1/3}
\end{equation}
\noindent We have plotted in figure 6 the variations of three
functions $C^{(l,s)}(T^{\ast})$. We are mostly interested in their
low-temperature limits, which are necessary for our calculations.
We thus obtain $C^{(2,2)}(0) = 3.00$, in excellent agreement with
the value $2.99$ obtained by BV \cite{beijerinck81}, $C^{(1,2)}(0)
=3.94 $ and $C^{(1,1)}(0)= 1.48$. We also get the value of the
ratio $\rho_o = C^{(1,2)}(0)/C^{(2,2)}(0) = 1.312\pm 0.004$, also
in very good agrement with the value $\rho_o = 1.32$ deduced from
the analytic deflection function of BV \cite{beijerinck81}.

\section{Appendix B: calculation of ${\mathbf F}_{\|,2}$ and ${\mathbf F}_{\bot,2}$}

We must calculate:

\begin{equation}
\label{B1} {\mathcal F}_{X,2}\approx \frac{n_1}{k_B u_{\infty}}
\int \langle g \frac{d\sigma(g)}{d\Omega}\rangle d\Omega \int
\Delta E_{X,2} f_1({\mathbf v}_1) f_2({\mathbf v}_2) d^3{\mathbf
v}_1 d^3 {\mathbf v}_2 d\Omega
\end{equation}

\noindent with $X= \|$ and $\bot$. We introduce the center of mass
velocity ${\mathbf v}_{cm}$

\begin{equation}
\label{B2} {\mathbf v}_{cm} = \alpha {\mathbf v}_1 +\beta {\mathbf
v}_2
\end{equation}

\noindent with $\alpha = m_1/M$, $\beta = m_2/M$ and $M=m_1 +
m_2$. The relative velocity before the collision is ${\mathbf g}
={\mathbf v}_1 - {\mathbf v}_2$ and after the collision is
${\mathbf g}_f$. The deflection angle noted $\chi$ is the angle
between ${\mathbf g}$ and ${\mathbf g}_f$, which have the same
modulus. We must also introduce the azimuth $\varphi$ measuring
the orientation in space of the plane ${\mathbf g}$, ${\mathbf
g}_f$. This azimuth $\varphi$ is equally probable in the $0\leq
\varphi <2\pi$ range. The final velocity of atom $2$ is given by:

\begin{equation}
\label{B4}{\mathbf v}_{2f} = \alpha {\mathbf v}_1 +\beta {\mathbf
v}_2 -\alpha {\mathbf g}_f
\end{equation}
\noindent We express its parallel and perpendicular components as
a function of the initial velocities, of the relative velocity and
of the various angles and we want to calculate:

\begin{equation}
\label{B5}\Delta E_{2X} = \frac{m_2}{2} \left({\mathbf v}_{2fX}^2
-{\mathbf v}_{2X}^2\right)
\end{equation}

\noindent We then integrate these quantities over the angles and
over the velocities ${\mathbf v}_1$ and ${\mathbf v}_2$ with the
normalized functions $f_1({\mathbf v}_1) f_2({\mathbf v}_2)$. All
the terms involve double products of the type $v_{iX}v_{jY}$,
where $i,j$ stands for atom indices $1,2$ and $X,Y$ for $\|$ or
$\bot$. After integration, all these products vanish, excepted the
ones with $i=j$ and $X=Y$ and their values are simply related to
the temperature associated to this atom and this degree of
freedom:
\begin{eqnarray}
\label{B6}\Delta E_{2\|} &=& \frac{m_2}{2} \left[
\frac{k_BT_{\|1}}{m_1} \alpha^2 \left(1- \cos\chi\right)^2 -
\frac{k_BT_{\|2}}{m_2} \left(\alpha^2 \sin^2\chi
+2\alpha\beta\left(1-\cos\chi\right)\right) \right. \nonumber \\
&+& \left. \left(\frac{k_BT_{\bot1}}{m_1}+\frac{k_BT_{\bot2}}{m_2}
\right) \alpha^2 \sin^2\chi\right]
\end{eqnarray}

\begin{eqnarray}
\label{B7}\Delta E_{2\bot} &=& \frac{m_2}{2} \left[\left(
\frac{k_BT_{\|1}}{m_1} + \frac{k_BT_{\|2}}{m_2} \right)\alpha^2
\sin^2\chi + \frac{k_BT_{\bot1}}{m_1} \alpha^2 \left(3 + \cos^2
\chi - 4\cos \chi)\right) \right. \nonumber
\\ &-& \left. \frac{k_BT_{\bot2}}{m_2} \left( 4 \alpha \beta \left(1- \cos\chi\right)
+ \alpha^2 \sin^2\chi \right)\right]
\end{eqnarray}
\noindent Finally we integrate over $\chi$. The integrals involve
the angular weights $(1- \cos^l\chi)$ with $l=1$ and $2$ only,
i.e. they involve only $Q_{1,2}^{(1)}$ and $Q_{1,2}^{(2)}$ defined
by equation (\ref{A1}). There was no terms involving $l=1$ terms
in the pure gas case. This new feature is due to the fact that the
forward-backward symmetry existing in the pure gas case is now
broken by the different masses of the two colliding atoms and by
the fact that we calculate the energy transfer for atom $2$ only.

We have made an approximation by considering $g Q_{1,2}^{(l)}(g)$
as independent of $g$ and we must fix its value. To be coherent
with the pure gas case, we have made the same approximation in the
pure gas case and we have identified the two values of
$\Lambda(z)$:
\begin{equation}
\label{B8} \langle g Q_{1,2}^{(l)}(g)\rangle = \frac{32}{15}
\Omega_{1,2}^{(l,2)}(T_m)
\end{equation}
\noindent As $\Omega_{1,2}^{(l,2)}(T_m)$ varies very slowly with
$T_m$, like $T_m^{1/6}$, the exact choice for $T_m$ is not
important and we have always used $T_m = (T_{\|1} + 2
T_{\bot1})/3$.

%%%%%%%%%%%%%%%%%%%%%%%%%%%%%%%%%%%%%%%%%%%%%%%%%%%%%%%%%%%%%%%%%%%

%%%%%%%%%%%%%%%%%%%%%%%%%%%%%%%%%%%%%%%%%%%%%%%%%%%%%%%%%%%%%%%%%%%%%%
%%%%%%%%%%%%%%%%%%%%%%%%%%%%%%%%%%%%%%%%%%%%%%%%%%%%%%%%%%%%%%%%%%%

\newpage
\begin{figure}
  \caption{The reduced temperatures $\tau_{Xi}$ are plotted as a
function of the reduced z-coordinate $\zeta$. For $\tau_{\|1}$ and
$\tau_{\bot1}$, this plot is identical to figure 5 of Beijerinck
and Verster $[5]$. The ultra cooling appears on $\tau_{\|2}$,
which reaches a terminal value lower than $\tau_{\|1}$. The
parallel and perpendicular temperatures of the seeded gas separate
for larger $\zeta$ values than the temperatures of the carrier gas
and this fact is the basis of the ultra cooling of the seeded gas.
The present calculation corresponds to the case of lithium seeded
in argon, with $\rho_s = 2.55$ and $\rho_o = 1.32$, and it
predicts $T_{\|2\infty}/T_{\|1\infty} = 0.38$.}
\end{figure}

\begin{figure}
\caption{The calculated parallel temperature ratio
$T_{\|2\infty}/T_{\|1\infty}$ is plotted as a function of the mass
ratio $m_2/m_1$, spanning the range from hydrogen to rubidium
seeded in argon. Each curve corresponds to a different value of
the ratio $\rho_s$ and is labeled by this value.}
\end{figure}

\begin{figure}
    \caption{\label{fluorescence} Laser induced fluorescence signal as a function of the
   laser frequency: the dots represent the experimental data and the full curves are the Gaussian fits to
   each line. The Fabry-Perot used for calibration has a free
   spectral range equal to $251.4 \pm 0.5$ MHz. The saturated
   absorption spectrum providing lithium Doppler-free peaks is
   represented for frequencies above $1250$ MHz. The fluorescence
   peaks are labeled by the ground state $F$ value ($F = 1$ or $2$)
   and by a letter corresponding to the laser beam. The angle between
   the atomic beam and the laser beam is $\theta_A =47.9\pm
   0.5^\circ$ for beam A and $\theta_B\approx 90^\circ$ for beam B.}
\end{figure}

\begin{figure}
  \caption{\label{pargon} The parallel temperature of lithium
seeded in argon as a function of argon pressure in the source. The
points represent our measurements. The full curve is the result of
our theoretical calculation with $T_{\|2} /T_{\|1} =0.38$. The
agreement between theory and experiment is quite good. The dashed
curve results from a fit of the four data points with an argon
pressure below $400$ millibar: the best fit corresponds to a
temperature ratio $T_{\|2}/T_{\|1} =0.31$. For larger source
pressures, the lithium parallel temperature remains roughly
constant, in disagreement with theory (see the discussion in the
text).}
\end{figure}

\begin{figure}
  \caption{\label{ylg} Plot of the variations of
   $Y^{(l)}=Q^{(l)}(g) E^{\ast 1/3}/\sigma^2$ as a function of
   $E^{\ast}$ for $l=1$ (upper curve) and $l=2$ (lower curve). Both
   quantities approach their limit rapidly when $E^{\ast}<0.5$.}
\end{figure}

\begin{figure}
    \caption{\label{cls} Plot of the variations of the quantities
   $C^{(l,s)}(T^{\ast})$ defined by equation (\ref{A6}) as a function
   of $T^{\ast}= k_BT /\epsilon$ for $(l,s) = (1,1)$ (dots); $(1,2)$ (dash-dotted curve) and
   $(2,2)$ (solid curve).}
\end{figure}

%%%%%%%%%%%%%%%%%%%%%%%%%%%%%%%%%%%%%%%%%%%%%%%%%%%%%%%%%%%%%%%%%%%

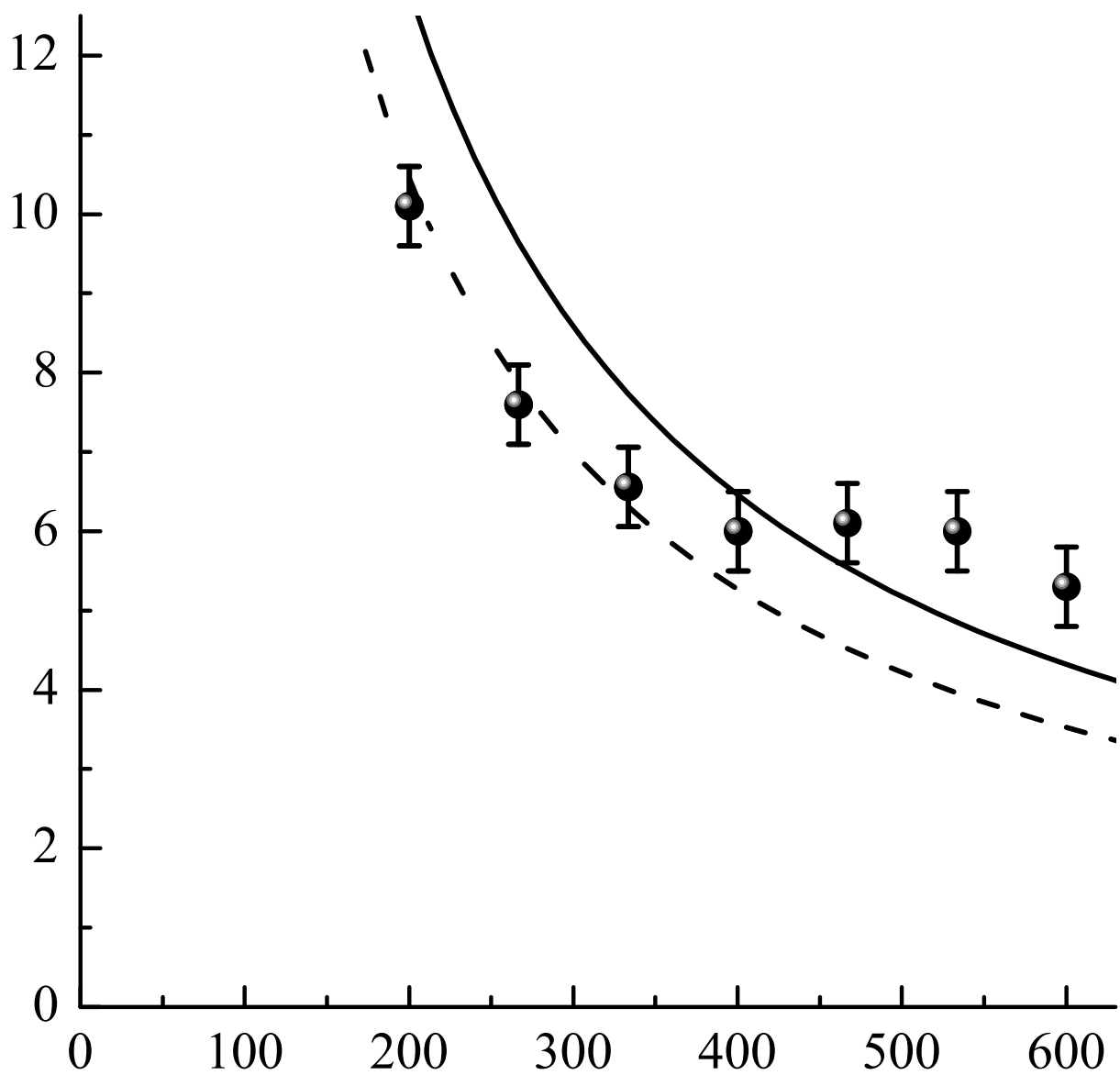
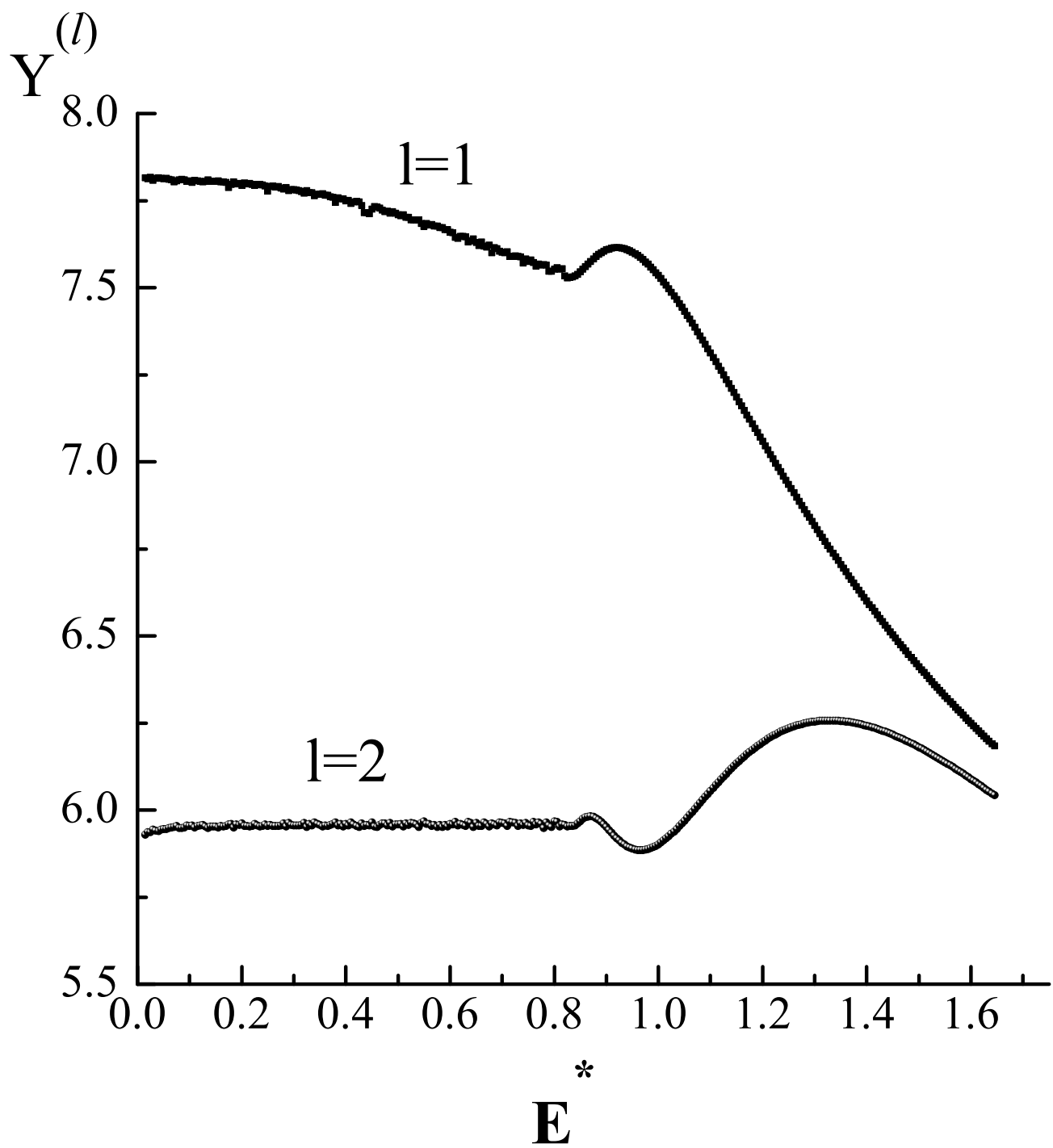
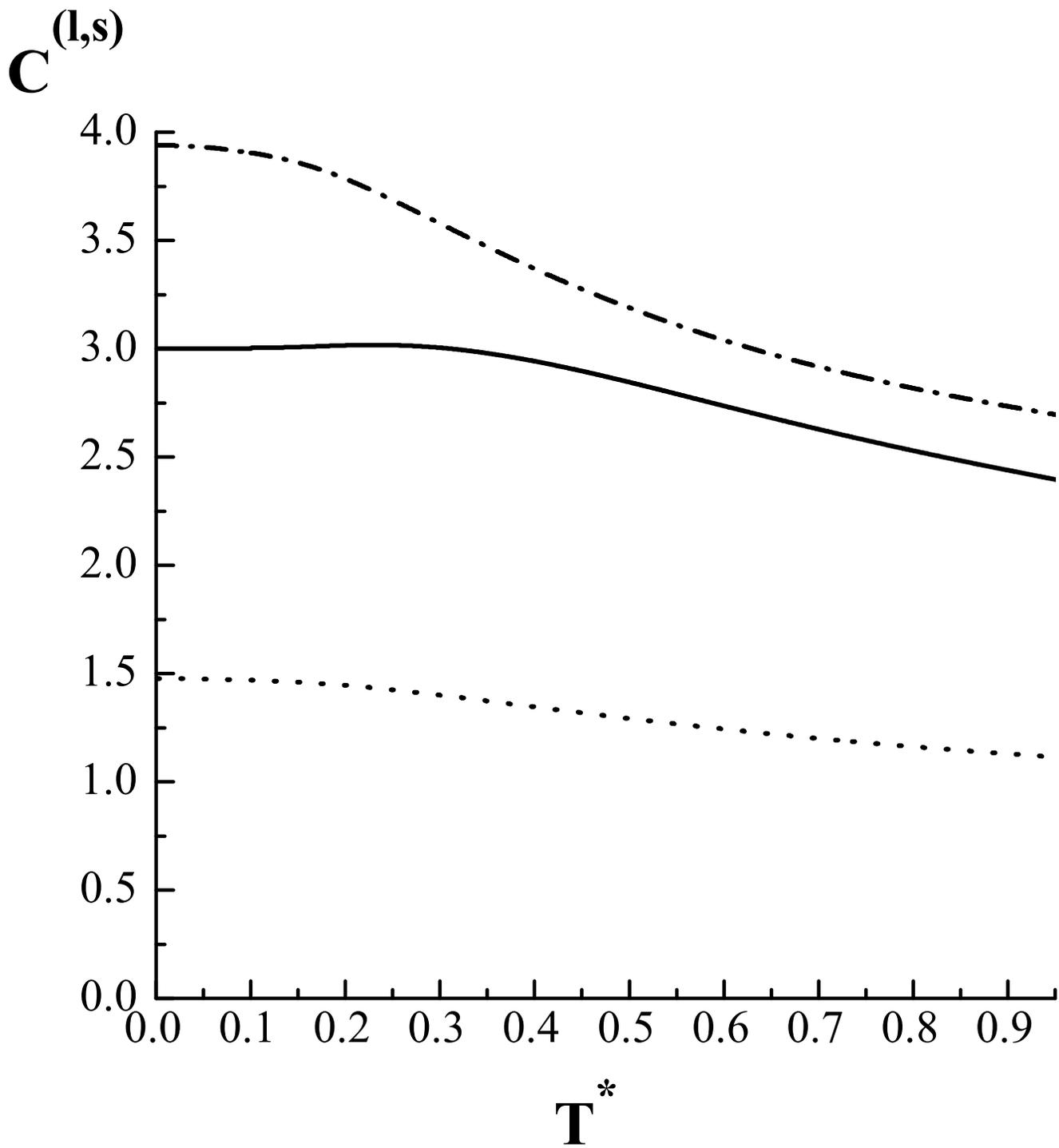
\end{document}